\documentstyle[12pt]{article}
\begin{document}
\baselineskip=0.84cm
\topmargin -0.5in
\vspace{0.5in}
\rightline{MAD/TH/92-8\qquad}
\medskip
\begin{center}
       {\large\bf Jets and jet multiplicities
    in high energy photon-nucleon interactions} \\
\bigskip
{ Loyal Durand and Katsuhiko Honjo}\\
    {\it Department of Physics, University of Wisconsin-Madison,\\
        Madison, Wisconsin 53706} \\
{Raj Gandhi}\\
{\it Department of Physics, Texas A\&M University,\\
College Station, TX 77843}\\
{Hong Pi}\\
{\it Department of Theoretical Physics, University of Lund,\\
Lund, Sweden S-22362}\\
{Ina Sarcevic}\\
{\it Department of Physics, University of Arizona,\\
Tucson, Arizona 85721}\\
\end{center}
\medskip
\begin{center}
Abstract
\end{center}

We discuss the theory of jet events in high-energy
photon-proton interactions using a model which gives a good description
of the data available
on total inelastic $\gamma p$ cross sections up to $\sqrt{s}$=210 GeV.
We show how to calculate the jet cross sections and jet multiplicities
and give predictions for these quantities for energies appropriate for
experiments at the HERA
$ep$ collider and for very high energy cosmic ray
observations.
\medskip

\leftline{PACS number(s): 13.60.Hb, 13.87.Ce, 12.40.Lk}
\medskip

The hadronic structure of the photon and its interaction with hadrons are
currently subjects of considerable interest.The ZEUS and H1 groups at the
{\it ep} collider HERA have presented
preliminary results \cite{hera} on the $\gamma p$ photoproduction
cross section at 210 GeV center-of-mass energy which show
the rise with energy typical of other hadronic interactions. It was shown
by Gandhi and Sarcevic \cite{raj} in a simple model that this rise can
be used to discriminate between different sets of photon structure functions.
The hadronic behavior of the photon
at very high energies ($\sim 10^{8}$ GeV photon energy) is
also of interest for the theory of
air showers triggered by cosmic ray photons \cite{airshowers}.

We have developed an improved QCD-based model
of the photon interaction with
nucleons \cite{honjo} which gives predictions for the total inelastic
$\gamma p$ cross section which agree well with the HERA data. The model has
the interesting feature that the information needed to calculate
$\sigma_{inel}^{\gamma p}$ is largely determined by high energy $\pi N$
scattering. In this paper, we will use the model to discuss
the total jet cross section $\sigma_{jet}^{\gamma p}$ and the
probabilities for multiple-jet events,
and give our predictions for those quantities at HERA energies
and the energies relevant for cosmic ray experiments.

We begin by sketching our model for inelastic $\gamma p$ scattering. The
theory is developed in more detail in \cite{honjo}.
The incoming physical photon state $\vert \gamma \rangle_{phys}$
can be expressed as a superposition of the bare photon and the set of virtual
hadronic states $\vert m \rangle$ which can be distinguished in the subsequent
interaction, and appear with probabilities ${\cal P}_{m}$.
Accordingly, the total inelastic $\gamma p$ cross section
is given in a semiclassical picture as
\begin{equation}
            \sigma^{\gamma p}_{inel} = (1 - {\cal P}_{had})\sigma_{dir}
                                  + \sum_{m}\,\!'\,{\cal P}_{m}\, \sigma^{mp},
            \hspace{1cm} {\cal P}_{had}= \sum_{m}\,{\cal P}_{m}\ll 1.
\end{equation}
Included among the states $\vert m \rangle$ are low-mass vector meson states
such as the $\rho, \omega$ and $\phi$, and complex non-resonant final states
which can be described on the average in a quark-gluon basis.

Whether the photon appears as a virtual
vector meson state or a high-mass non-resonant state depends on the relative
transverse momentum $2p_{\perp 0}$ of the quark and antiquark in the virtual
transition $\gamma\rightarrow q\overline{q}$ which initiates the hadronic
interaction. We note that $p_{\perp 0} \approx
1/r_{\perp}$, where $r_{\perp}$ is the average transverse separation of the
virtual quarks during the lifetime of the $q\overline{q}$ system.
If $r_{\perp}$ is greater than, or on the order of, the
average transverse radius $R_{\perp}\equiv 1/Q_0$ of a vector meson,
QCD confinement effects
will clearly set in, and the $q\overline{q}$ system will most likely
appear in a hadronic collision as a light vector meson.
For $r_{\perp}<R_{\perp}$ the $q\overline{q}$ system
will be smaller than a vector meson.
In this picture, the hadronic system behaves for $p_{\perp 0}<Q_0$
like a vector meson, and for $p_{\perp 0}>Q_{0}$, like
a system of quasi-free quarks and gluons with a transverse area smaller than
that of a vector meson by a factor $(Q_{0}/p_{\perp 0})^{2}$. Its
interaction cross section will be reduced accordingly.

After eikonalizing the hadronic cross sections using an
impact parameter representation to
account for possible multiple parton scatterings
in a single $\gamma p$ collision,
the  total inelastic $\gamma p$ cross section is given by \cite{honjo} as
\begin{eqnarray}
   {\displaystyle \sigma^{\gamma p}_{inel}}
    &=&{\displaystyle\sigma_{dir}
       +\lambda {\cal P}_{\rho}
        \int d^{2}b\, (1-e^{\textstyle -2Re\chi^{\rho p}})}  \nonumber\\
    & &{\displaystyle +\sum_{q}e_{q}^{2} \frac{\alpha_{em}}{\pi}
        \int_{Q_{0}^{2}}\frac{dp_{\perp 0}^{2}}{p_{\perp 0}^{2}} \int d^{2}b\,
        (1-e^{\textstyle -2Re\chi^{q\overline{q} p}})}.
\end{eqnarray}
Here ${\cal P}_{\rho}$ is the probability that a photon exists
in the $\rho$ meson
state, ${\cal P}_{\rho}= 4\pi\alpha_{em}/f_{\rho}^{2} $,
where $f_{\rho}^{2}$
is the $\gamma \rho$ coupling in the vector meson dominance model.
We will treat the $\rho, \omega$ and $\phi$ as equivalent, and will use the
quark model ratios for the couplings $f_{V}$; then $\lambda = 4/3$ for equal
$\rho p,  \omega p$, and $\phi p$ cross sections, and $\lambda = 10/9$ for
complete suppression of the $\phi$ contribution at low $p_{\perp 0}$
\cite{phi}.
The third term in Eq.~(2) includes
the contributions from the excited hadronic states of the photon,
which we collectively call $q\overline{q}$ states.
The factor $(e_{q}^{2}\alpha_{em}/\pi p_{\perp 0}^{2})dp_{\perp 0}^{2}$
is just the differential probability of producing
a $q\overline{q}$ pair at a relative transverse momentum $2p_{\perp 0}$.

The real part of the eikonal function for the $\rho p$
interaction can be written in the form \cite{durand}
\begin{eqnarray}
   Re\chi^{\rho p}(b,s)
         &=&Re\chi^{\rho p}_{soft}(s) + Re\chi^{\rho p}_{QCD}(b,s) \nonumber\\
         &=&\frac{1}{2} A^{\rho p}(b)
            [\sigma_{soft}^{\rho p} (s)+ \sigma_{QCD}^{\rho p} (s)],
\end{eqnarray}
where $A^{\rho p}(b)$ is the density overlap function,
\begin{equation}
       A^{\rho p} (b) = \int d^{2}b'\rho_{\rho}(b)\rho_{p}(|\vec{b}-\vec{b'}|),
       \hspace{1cm} \int d^{2}b A^{\rho p}(b) = 1,
\end{equation}
and $\sigma_{soft}^{\rho p}$ and $\sigma_{QCD}^{\rho p}$ are the soft- and
hard-scattering parts of the intrinsic cross section.
$\sigma_{soft}^{\rho p}$ was  parametrized in \cite{honjo} using a Regge-like
form,
\begin{equation}
    \sigma_{soft}^{\rho p} = \sigma_{0}+\sigma_{1}(s-m_{p}^{2})^{-1/2}+
                                        \sigma_{2}(s-m_{p}^{2})^{-1},
\end{equation}
while $\sigma_{QCD}^{\rho p}$ was identified with
the inclusive parton-level cross section for $\rho p$ scattering,
\begin{equation}
    \sigma_{QCD}^{\rho p}=\sum_{ij}\frac{1}{1+\delta_{ij}}\int_{0}^{1}dx_{1}
                          \int_{0}^{1}dx_{2}\int_{p_{\perp,min}^{2}}dp_{\perp}
                          ^{2}f_{i}^{p}(x_{1},p_{\perp}^{2})f_{j}^{\rho}
                          (x_{2},p_{\perp}^{2})\frac{d\hat{\sigma}_{ij}}
                          {dp_{\perp}^{2}}.
\end{equation}

Using the equivalence of the $\rho$ and $\pi$ states
in the quark model, we can equate the parton distribution and
density profile functions for the $\rho$ meson to the corresponding
functions for the pion and take $f^{\rho}_i$=$f^{\pi}_i$ and $\rho_{\rho}
(b)$=$\rho_{\pi}(b)$. The parton distributions in the pion are known
reasonably well \cite{owens}.
The density profile functions $\rho_{p}(b)$ and $\rho_{\pi}(b)$ can be
taken as the Fourier transforms of the electromagnetic form factors of the
proton and the $\pi$ meson.  The density overlap function $A^{\rho p}\approx
A^{\pi p}$ is then given by \cite{pion}
\begin{equation}
  A^{\rho p}(b)=\frac{1}{4\pi}\frac{\nu^{2}\mu^{2}}{\mu^{2}-\nu^{2}}
  \left\{\nu bK_{1}(\nu b)-\frac{2\nu^{2}}{\mu^{2}-\nu^{2}}[K_{0}(\nu b)-K_{0}
  (\mu b) ]\right\},
\end{equation}
where $K_{n}(x)$ is a hyperbolic Bessel function, and
the ``size'' parameters of the pion and the proton are
$\mu^{2}=$ 0.47 GeV$^{2}$ and $\nu^{2}=$ 0.71 GeV$^{2}$, respectively.
The former corresponds to a root-mean-square transverse radius for the
pion or $\rho$ meson $R_{\perp}$=$1/Q_0\approx2/\mu$.

The eikonal function for the $q\overline {q}p$ scattering terms in Eq.~(2)
can be expressed similarly as
\begin{equation}
    2Re\chi^{q\overline{q} p} =
         A^{q\overline{q} p}(b,p_{\perp 0})
         [\sigma_{soft}^{q\overline{q} p}(s,p_{\perp 0})
         +\sigma_{QCD}^{q\overline{q} p}(s,p_{\perp 0})].
\end{equation}
We assume  that $\rho_{q\overline{q}}(b)$ has the same form
as $\rho_{\rho}(b)$
except with the size parameter $\mu=2Q_0$ of the $\rho$ meson
replaced by $2p_{\perp 0}$ \cite{honjo}. With this replacement,
$A^{q\overline{q} p}(b,p_{\perp 0})$
and $\rho_{q\overline{q}}(b,p_{\perp 0})$ are continuous with the corresponding
functions for the $\rho$ meson at $p_{\perp 0}=Q_{0}$, but describe a shrinking
system for $p_{\perp 0}>Q_0$.
Finally, the cross sections $\sigma_{soft}^{q\overline{q} p}(s,p_{\perp 0})$
and $\sigma_{QCD}^{q\overline{q} p}(s,p_{\perp p})$,
were assumed to scale with the
physical cross sectional area of the $q\overline{q}$ system so that
\begin{equation}
   \sigma_{soft}^{q\overline{q} p}(s,p_{\perp 0})
   +\sigma_{QCD}^{q\overline{q} p}(s,p_{\perp 0})  \nonumber \\
          = (Q_{0}^{2}/p_{\perp 0}^{2})
            [\sigma_{soft}^{\rho p}(s)+\sigma_{QCD}^{\rho p}(s)].
\end{equation}
This model guarantees that $\sigma_{soft}$ and $\sigma_{QCD}$
are continuous at $p_{\perp 0}=Q_{0}$, and that the soft contributions from
the $q\overline{q}$ states die out as $1/p_{\perp 0}^2$ as expected
for higher twist contributions.

We have calculated the total inelastic $\gamma p$ cross section
using Eq.~(2) and the assumptions above. The soft cross section in Eq.~(5)
was determined \cite{honjo} using the low-energy $\gamma p$ data
\cite{lowenergy}.  The calculation used the parton
distributions of Owens \cite{owens} for the pion and those of Eichten
{\it et al.} \cite{ehlq} for the proton. The parameter $p_{\perp ,min}$
in Eq.~(6) which determines the transverse momentum at which hard
parton-level collisions come into play was taken from a similar fit to
$\pi^{\pm} p$ scattering \cite{pion}. The results agree
with the preliminary HERA data \cite{hera} and those at lower energies
as shown by the top curve in Fig.~1(a).
The calculated rise in $\sigma^{\gamma p}_
{inel}$ at higher energies arises from the hard-scattering contributions, and
is a prediction of the model rather than a fit to the HERA data. It is
therefore of interest to look for other tests of the model. The jet cross
sections predicted by the model provide one such test.

The total jet cross section $\sigma_{jet}^{\gamma p}(s,Q^{2})$ is defined to
be the part of the inelastic $\gamma p$ cross section which includes events
with at least one semihard parton-parton\,
(or $\gamma$-parton) scattering with
a momentum transfer $p_{\perp}^{2} \geq Q^{2}$, irrespective of any soft
processes that may occur.  To find an expression for
the jet cross section $\sigma_{jet}^{mp}$ in the interaction between the
proton and the hadronic state $\vert m \rangle$ of the photon,
we use the fact that semiclassically
$exp{\textstyle [-2Re\chi_{QCD}}(b,s,Q^{2})]$ can
be interpreted as the probability that there is {\it no} parton-parton
scattering with $p_{\perp}^{2} \geq Q^{2}$ in a hadronic collision at impact
parameter $b$.  Using this observation, we can rewrite the expression for
the total hadronic cross section $\sigma^{mp}$ to separate out the jet-free
part,
\begin{eqnarray}
  \sigma_{had}^{mp}(s)
  &=&\sigma_{nojet}^{mp}(s,Q^{2})+\sigma_{jet}^{mp}(s,Q^{2})\nonumber \\
  &=&\int d^{2}b\left(1-e^{\textstyle -2Re\chi_{QCD}^{mp}
                                      -2Re\chi_{soft}^{mp}}\right) \nonumber \\
  &=&\int d^{2}b\left(1-e^{\textstyle -2Re{\chi_{soft}^{mp}}'(b,s,Q^{2})}
         \right)e^{\textstyle -2Re\chi_{QCD}^{mp}(b,s,Q^{2})}     \nonumber\\
  & &+\int d^{2}b\left(1-e^{\textstyle -2Re\chi_{QCD}^{mp}(b,s,Q^{2})}\right).
\end{eqnarray}
Here
\begin{equation}
     \sigma_{jet}^{mp}(s,Q^{2})=\int d^{2}b
      \left(1-e^{\textstyle -2Re\chi_{QCD}^{mp}(b,s,Q^{2})}\right)
\end{equation}
is the total cross section for events associated with $\vert m\rangle$ which
contain jets with $p_{\perp}\ge~Q$.
The factor $\sigma^{\gamma p}_{QCD}$ which appears in the
eikonal function in Eq.~(11) is now to be evaluated using the expression in
Eq.~(6) with $p_{\perp ,min}^2$ replaced by $Q^2$.
The remaining ``nojet'' cross section involves a modified
eikonal function
\begin{equation}
     Re{\chi_{soft}^{mp}}'(s,b,Q^{2})=Re\chi_{soft}^{mp}+Re\chi_{QCD}^{mp}
                                    -Re\chi_{QCD}^{mp}(b,s,Q^{2}),
\end{equation}
and includes contributions from ``soft'' jets
with $p_{\perp,min}^{2} \leq p_{\perp}^{2} \leq Q^{2}$ as well as the usual
soft term.

The total jet cross section in $\gamma p$ scattering is the generalized
sum of the cross sections $\sigma_{jet}^{mp}(s,Q^{2})$,
weighted with the probabilities
${\cal P}_{m}$, over all the possible hadronic states of the photon, plus the
very small jet contribution from
$\sigma_{dir}(s,Q^{2})$,
\begin{eqnarray}
 \sigma_{jet}^{\gamma p}(s,Q^{2})
 &=&\sigma_{dir}(s,Q^{2})+{\displaystyle \sum_{m}{\cal P}_{m}\sigma_{jet}^{m}}
    \nonumber\\
 &=&\sigma_{dir}(s,Q^{2})+\lambda\,{\cal P}_{\rho}\,
    \sigma_{jet}^{\rho p}(s,Q^{2})   \nonumber  \\
 & &+{\displaystyle \sum_{q}e_{q}^{2}\,\frac{\alpha_{em}}{\pi}
    \int_{Q_{0}^{2}}\frac{dp_{\perp 0}^{2}}{p_{\perp 0}^{2}}\,\sigma_{jet}
    ^{q\overline{q} p}(s,Q^{2},p_{\perp 0}^{2}) }.
\end{eqnarray}
The jet cross sections for specific kinematic cuts, e.g., on jet angle,
can be calculated using the same basic formulas with the cuts imposed
on the integral for $\sigma_{QCD}^{\gamma p}$ in Eq.~(6).

We have calculated the total jet cross section $\sigma_{jet}^{\gamma p}$
in the HERA
energy range for $Q$ = 2, 3, 4, and 5 GeV, using the same parameters as were
used in \cite{honjo}. Our predictions are shown in Fig.~1 along with the
calculated total inelastic  $\gamma p$ cross section
and the experimental data \cite{hera,lowenergy}.
At $\sqrt s$=200 GeV, the jet cross section is predicted to be approximately
1.3\% of the total inelastic cross section for $Q$= 5 GeV in rough
agreement with earlier predictions \cite{raj}, and 25\% of the
total for $Q$= 2 GeV.

The rapid growth of the jet cross section with energy is associated
with the growth of the parton distribution functions in the photon and
the proton at the small values of $x$ which become accessible at high
energies. This growth also leads to an increasing probability of the
scattering of
more than one parton in a single $\gamma p$ collision.For this reason,
the inclusive parton-level $\gamma p$ cross section $\sigma^{\gamma p}_{QCD}$,
which counts each parton collision separately, is not the same as $\sigma^
{\gamma p}_{jet}$. The latter has the overcounting due to multiple independent
scatterings removed in the eikonalization.
We show the effect of the eikonalization up to HERA energies in Fig.~2.
The effect is not very significant for
$Q$  = 3, 4 and 5~GeV at $\sqrt s$= 200 GeV,
but causes about a 10\% reduction in the jet cross section for $Q$  = 2~GeV.

In Fig.~2 we show $\sigma_{jet}^{\gamma p}$ for the cosmic ray energy
range for the same choices of $Q$  as in Fig.~1,
together with the calculated cross sections
$\sigma_{inel}^{\gamma p}$ and $\sigma_{QCD}^{\gamma p}$ for $Q$  = 2 GeV.
At very
high energies, the jet cross section clearly comprises a large part of the
total inelastic cross section,
e.g., at $E_{\gamma}=2\times 10^{8}$ GeV
(or $\sqrt{s}\approx 2\times 10^{4}$ GeV)
$\sigma_{jet}^{\gamma p}$ constitutes $\approx 75\%$
of $\sigma_{inel}^{\gamma p}$ for $Q$  = 2 GeV.
Even for $Q$  = 5 GeV, $\sigma_{jet}^{\gamma p}$ gives nearly 40\% of
$\sigma_{inel}^{\gamma p}$.
Also, as more of the incident partons scatter,
the effect of eikonalization becomes quite large as shown for $Q$= 2 GeV.

To calculate the multiplicity of jets, we note that the average
number of parton-parton scatterings with $p_{\perp}>Q$ in a single
$\gamma p$ collision at impact parameter b associated with the hadronic
component $\vert m\rangle$ of the photon is
\begin{equation}
     \bar{n}_{m}(b,s,Q^{2})=\sigma_{QCD}^{mp}(s,Q^{2})A^{mp}(b)
                           =2Re\chi_{QCD}^{mp}(b,s,Q^{2}).
\end{equation}
Since the parton-parton scatterings in our model are independent,
the probability of having n scatterings (2n jets) in such a hadronic collision
has a Poisson distribution,
\begin{equation}
     P_{n}^{mp}(b,s,Q^{2})=\frac{1}{n!}[\bar{n}_{m}(b,s,Q^{2})]^{\textstyle n}
                           e^{\textstyle -\bar{n}_{m}(b,s,Q^{2})}.
\end{equation}
The probability $P_{n}(s,Q^{2})$ of having n scatterings
relative to {\it all} the inelastic events at an impact parameter b is then
\begin{eqnarray}
     P_{n}(s,Q^{2})
         &=&{\displaystyle (\sigma_{inel}^{\gamma p})^{-1} \sum_{m}{\cal P}_{m}
             \sigma_{jet,n}^{mp}}  \nonumber\\
         &=&{\displaystyle (\sigma_{inel}^{\gamma p})^{-1} \sum_{m}{\cal P}_{m}
              \int d^{2}b\, P_{n}^{mp}(b,s,Q^{2}) }  \nonumber\\
         &=&{\displaystyle (\sigma_{inel}^{\gamma p})^{-1}
             \sum_{m}{\cal P}_{m} \frac{1}{n!}
             \int d^{2}b [\bar{n}_{m}(b,s,Q^{2})]^{\textstyle n}
             e^{\textstyle -\bar{n}_{m}(b,s,Q^{2})}}
\end{eqnarray}
for $n\ge 1$. $P_{0}(s,Q^{2})$ is defined to be the
probability of having a soft interaction with no associated jets,
\begin{equation}
     P_{0}(s,Q^{2})=(\sigma_{inel}^{\gamma p})^{-1}
               \sum_{m}{\cal P}_{m} \int d^{2}b
               \left(1-e^{\textstyle -2Re{\chi_{soft}^{mp}}'(b,s,Q^{2})}\right)
               e^{\textstyle -\bar{n}_{m}(b,s,Q^{2})}.
\end{equation}
Note that we have neglected the very small two-jet contribution to
these expressions from the direct QCD
interaction of the photon with the partons in the proton.

We have calculated  $P_{n}$ for n = 0, 1, 2 and 3
at $\sqrt s $= 200 GeV using $Q$  = 2, 3, 4, and 5 GeV. The results are shown
in Table I. For $Q$  = 5 GeV the probability of having a 2-jet
event (a single parton scattering) is about 1\%;
the probability for 2n-jet events with n larger than 1
is essentially zero.  For $Q$  = 2 GeV the {\em total} probability of having
any
hadronic jet event is approximately 27\%, with the 2-jet events comprising
more than 90 \% of the total.

We also investigated jet production in $\gamma p$ scattering at
very high energies. In Fig.~3 we show
the distribution of $P_{n}$ at $\sqrt{s}=2\times 10^{4}$ GeV
with the choice of $Q$  = 3~GeV.  It is more likely at this energy to have
jets in an event than not, with an approximately 70\% jet probability.
While the 2-jet production is still the most probable,
events with more than 2 jets occur with about 20\% probability.

It will be quite interesting to see if the predictions above are verified in
future experiments since they connect the hadronic properties of the photon
quite directly to those of the pion. In particular, the jet structure of pion
and photon-induced reactions on protons should be quite similar.

\medskip
\begin{center}
{ACKNOWLEDGMENTS}
\end{center}
This work was supported in part through U.~S. Department of Energy
Grants Nos.~DE-AC02-76ER00881 and DE-FG02-85ER40213. One of the authors (RG)
would like to thank the World Laboratory for support.

\begin{center}
FIGURE CAPTIONS
\end{center}

FIG.~1. (a) The calculated total jet cross sections as functions of $\sqrt{s}$
and the transverse momentum cutoff $Q$. The total $\gamma p$ inelastic cross
section calculated in \cite{honjo} is
compared to the low-energy \cite{lowenergy} and HERA \cite{hera} data
in the upper curve. (b) The eikonalized and inclusive jet cross sections for
the
same values of $Q$.

FIG.~2. The total inelastic and jet cross sections at cosmic ray energies.
$\sigma_{QCD}^{\gamma p}$ is the inclusive jet cross section.

FIG.~3. The probability distributions for $n$ parton-parton collisions or
$2n$ jets in a $\gamma p$ collision.

\newpage
\renewcommand{\arraystretch}{2}
\begin{table}[t]
 \begin{tabular*}{6in}[t]
       {@{\hspace{0.5cm}}@{\extracolsep{\fill}}ccccc@{\hspace{0.5cm}}}
 \hline \hline
$Q$ (GeV)  & $P_{0}$ & $P_{1}$ & $P_{2}$ & $P_{3}$   \\ \hline
2       & 0.734   & 0.228   & 0.020   & 0.002  \\
3       & 0.904   & 0.078   & 0.002   & 0.000  \\
4       & 0.954   & 0.030   & 0.000   & 0.000  \\
5       & 0.971   & 0.013   & 0.000   & 0.000  \\  \hline  \hline
 \end{tabular*}
 \caption{Probabilities $P_{n}$ of having $n$ parton-parton collisions
  or $2n$ jets in a single inelastic $\protect\gamma p$ event at
  $\protect\sqrt{s}$=200 GeV.}
\end{table}


\newpage
\begin{thebibliography}{20}
\bibitem{hera}The preliminary values of $\sigma_{inel}^{\gamma p}$ are:
ZEUS Collaboration, Phys.~Lett.~B {\bf 293}, 465 (1992),
$\sigma_{inel}^{\gamma p}$=154$\pm$16$\pm$32 $\mu$b at a average energy of
210 GeV;
H1 Collaboration (reported at the XXVI International Conference on High Energy
Physics by
F.~Eisele), $\sigma_{inel}^{\gamma p}$= 165$\pm$28 $\mu$b or 150$\pm$22 $\mu$b
depending on the model used in the analysis. We show a value 160$\pm$30 $\mu$b
for H1 in our figures to span the experimental and theoretical uncertainties.
\bibitem{raj} R.~Gandhi and I.~Sarcevic, Phys.~Rev.~D {\bf 44}, R10 (1991).
\bibitem{airshowers} See, for example, M. Drees and F. Halzen, Phys.~Rev.~Lett.
{\bf 61}, 275 (1988); M. Drees, F. Halzen, and K. Hikasa, Phys.~Rev.~D {\bf
39},
1310 (1989); T.K. Gaisser, F. Halzen, T. Stanev, and
E. Zas, Phys.~Lett.~B {\bf 243}, 444 (1990).
\bibitem{honjo} K. Honjo, L. Durand, R. Gandhi, I. Sarcevic, and H. Pi,
      University of Wisconsin Report No. MAD/TH/92-6.
\bibitem{phi} There is evidence for considerable suppression of the cross
section expected for photoproduction of the $\phi$ relative to that
 for the $\rho$ for $\sqrt{s}\approx$10 GeV. See R.M. Egloff {\it et al.},
Phys.~Rev.~Lett. {\bf 43}, 657 (1979). We therefore used $\lambda$=10/9 in the
calculations reported here. This has rather little effect on the high
energy results \cite{honjo}.
\bibitem{durand} L. Durand and H. Pi, Phys.~Rev.~Lett. {\bf 58}, 303 (1987);
Phys.~Rev.~D {\bf 40}, 1436 (1989).
\bibitem{owens}J.F. Owens, Phys.~Rev.~D {\bf 30}, 943 (1984).
\bibitem{pion} L. Durand and H. Pi, Phys.~Rev.~D {\bf 43}, 2125 (1991).
\bibitem{lowenergy} The lower energy data shown in Fig.~1 are from D.O.
Caldwell
{\it et al}., Phys.~Rev.~Lett. {\bf 25}, 609 (1970); {\it ibid.} {\bf 40}, 1222
(1978).
\bibitem{ehlq} E. Eichten, K. Lane, I. Hinchliffe, and C. Quigg, Rev.~Mod.~
        Phys.~{\bf 56}, 579 (1984).
\end{thebibliography}
\end{document}